# Observation of Kinetic Isotope Effect in Electrocatalysis with Fully Deuterated Ultrapure Electrolytes

Ken Sakaushi,[1]

[1]Center for Green Research on Energy and Environmental Materials, National Institute for Materials Science, 1-1 Namiki, Tsukuba, 305-0044 Ibaraki, Japan

**Corresponding to K.S.**: sakaushi.ken@nims.go.jp

**Abstract**

Kinetic isotope effect (KIE) is a common physicochemical effect to elucidate complicated microscopic reaction mechanism in biological, chemical and physical systems. Especially, the exchange of hydrogen to deuterium is a standard approach to investigate kinetics and pathways of a wide spectrum of key reactions involving proton transfer. However, KIE in electrocatalysis is still challenge. One main reason is owing to the high sensitivity to impurities in electrochemical systems. Aiming to establish an appropriate approach to observe KIE in electrocatalysis, we investigated KIE in electrocatalysis by using fully deuterated ultrapure electrolytes. With these electrolytes, we studied oxygen reduction reaction with platinum catalyst, which is well-known to be sensitive to impurity, as the model systems. In conclusion, the electrode processes in these systems can be strongly influenced by a purity of a selected deuterated electrolyte, especially in case of alkaline conditions. Therefore a highly pure deuterated electrolyte is indispensable to study microscopic electrode processes of electrocatalysis by analyzing KIE. This work shows a key criterion and methods to observe a reliable KIE in electrocatalytic systems, and therefore, provides a general approach to investigate complicated multielectron- and multiproton-transfer processes using not only standard electrochemical technique but also surface sensitive spectrometry.

# 1. INTRODUCTION

Mutilelectron-, mutilproton-transfer process is the core principle in a wide spectrum of energy transforming systems, such as metabolism in life or fuel-cells in green vehicles.[1-8] Therefore acquiring the microscopic picture of a variety of highly complicated multistep electrochemical processes can advance both fundamental scientific knowledge and modern technology. One of the simplest model systems to study the microscopic mechanism of aforementioned processes is the electrocatalysis at electrode/electrolyte interfaces using noble metals, such as platinum or gold.[5, 9-11] However, understanding microscopic electrocatalytic process at these well-known model systems is still a one of the today's challenges, even though there is a wide spectrum of advanced experimental and/or computational approaches to investigate surface reactions so far.[12-23] From this point, kinetic isotope effect (KIE) can be a powerful tool to clarify the basic reaction mechanism of electrocatalysis.[24-26] A use of heavy water in order to substitute proton in the system to deuteron (H-D exchange) is a classical topic in electrochemistry,[27-34] and this method was applied to analyze complicated electrode processes, such as electrocatalytic reactions on carbon-based electrocatalysts.[35-39] These non-noble metal based catalysts show a promising properties,[40-45] therefore is expected that KIE can be used for a part of approaches to find out the alternatives to present Pt-based electrocatalysts. However, a general method to observe/analyze KIE in electrochemical systems based on deuterated water was established recently.[46, 47] There are several reasons why it took such long time to reach to proper method to observe KIE in electrocatalysis. One reason is that thermodynamics in heavy water system is quite different compared to ordinary systems,[32, 46, 47] therefore we need modification to analytical equations for electrochemical reactions. Another reaction is that electrochemical systems are quite sensitive to impurity,[48-51] and this is one of main reasons for hindering to establish a procedure to ensure a reliable KIE measurement so far. Therefore a procedure shown here to provide the correct KIE observation in electrochemical system can give a huge impact.

In present study, we investigated an influence of impurities towards the KIE in oxygen reduction reaction (ORR) on a polycrystalline platinum (polyPt) electrode surface in acidic (0.05M $H_2SO_4$ and 0.05M $D_2SO_4$) and alkaline (0.1M KOH and 0.1M KOD) electrolytes. The particular focus lies in the identification of a method to observe correct KIE in electrochemical systems. Aiming to this goal, we prepared ultrapure deuterated water (denoted as Ultrapure) in order to perform the electrochemical experiments and compared with the results with the highly pure deuterated water, but not ultrapure grade, (denoted as Control). As the result, it was found that several key electrochemical features, such as electrochemically active surface area (ECSA), deuteron desorption isotherm, and Tafel slope, obtained in the Ultrapure systems are identical to the values obtained in the well-established ultrapure ordinary water systems. On the other hand, the different values and features are obtained in the Control deuterated electrolytes. This indicates that the Ultrapure-based systems can provide a sufficient clean condition to proceed a contamination-sensitive experiment. The kinetics of ORR electrode process were checked in order to confirm the impact of impurities towards the system by using a rotating ring-disk electrode technique (RRDE), and we found that the impurity can lead to kinetic values with a two or three order of magnitude difference. These different kinetic values can mislead to an unappropriated interpretation of KIE, and to incorrect microscopic picture of electrode processes. Therefore, a use of ultrapure heavy water is indispensable to obtain a reliable microscopic view of complicated multistep electrode processes.

## 2. EXPERIMENTAL METHODS

### 2.1. Standard Electrochemical Methods

The chemicals and equipment are the same with the previous reports,[46, 47] however, we described the details. The Electrochemical measurements were conducted using a RRDE set-up (Dynamic Electrode HR-301, HOKUTO DENKO) with an electrochemical analyzer (HZ-7000, HOKUTO DENKO) based on a custom-made three-compartment electrochemical glass cell at 298 K ± 1. The ring electrode was kept at 1.2 V *vs*. reversible deuterium electrode (RDeE). We use V vs. RDeE as the standard unit to show potential as $V_{RDeE}$ in this paper. We operate three different measurements and these data are summarized in order to show average values with errors. The cell was firstly cleaned by boiling in a mixture of concentrated sulfuric acid for overnight and then boiling in ultrapure ordinary water (MilliQ water, 18.3 MΩ cm) for overnight. Before electrochemical measurements, the cells were washed by an electrolyte (0.1M KOD in $D_2O$ or 0.05M $D_2SO_4$ in $D_2O$) several times, and then were poured the electrolyte to build systems. The electrolytes in the cell were bubbled with $O_2$ (purity > 99.999 %, Taiyo Nippon Sanso) or Ar (purity > 99.99995 %, Taiyo Nippon Sanso) for 30 min before the experiments to prepare the $O_2$-saturated condition or Ar-saturated condition, respectively. Resistance of electrochemical systems was measured prior to each experiment by using impedance measurements, and this value was used to correct *iR*-drop.

For preparation of electrolytes with ordinary water, a high purity KOH (semiconductor grade, 99.99% trace metals basis, Sigma-Aldrich) or $H_2SO_4$ (96 %, Ultrapur. grade, Merck) was mixed with ultrapure water (Milli-Q water, 18.3 MΩ cm). The electrolytes based on deuterated water were prepared by mixing potassium deuteroxide solution (40 wt. % in $D_2O$, Cambridge Isotope) or sulfuric acid-$d_2$ solution 96-98 wt. % in $D_2O$ (99.5 at. % D, Sigma-Aldrich), with high-purity heavy water ("100 %" distilled $D_2O$, Sigma-Aldrich) in order to obtain 0.1M KOD or 0.05M $D_2SO_4$ in $D_2O$. In order to prepare the ultrapure deuterated electrolytes, the as-received "100 %" $D_2O$ was purified by Simplicity UV (Merck KGaA, Darmstadt, Germany), to reach to the ultrapure quality (i.e. Milli-Q grade, 18.3 MΩ cm), and used it for preparing the fully deuterated electrolytes. The control electrolytes (denoted as Control) were prepared with the highly pure $D_2O$ ("100 %" distilled $D_2O$, Sigma-Aldrich) without additional purifications.

The three-electrode setup is consisted of a platinum counter electrode, a reversible deuterium electrode (RDeE) as the reference electrodes, and a commercial fixed polyPt working electrode (purchased from HOKUTO DENKOU) with a diameter of 0.5 cm (Therefore, the geometrical surface area of electrodes is 0.196 $cm^2$). Typically, we used a scan rate of 50 mV $s^{-1}$ and a rotation rate of 1600 rpm for the experiments. An electrochemically active surface area (ECSA) was measured by the typical method using cyclic voltammograms.[52] For example, 220 μC $cm^{-2}$ is assumed for a charge of full coverage of monolayer deuteron on a smooth polyPt surface.[53, 54] Therefore, a roughness factor (RF) of polyPt electrode and its ECSA can be obtained by the following equation:

RF = $Q_{D-UPD}$ (μC $cm^{-2}_{geo}$) / 220 (μC $cm^{-2}$), where μC $cm^{-2}_{geo}$ means a charge normalized by a geometrical surface area of an electrode,

ECSA ($cm^2_{ECSA}$) = RF × a geometrical surface area of an electrode (0.196 $cm^2$).

The RF of the electrode was measured in the ultrapure ordinary electrolytes, and applied this same electrode and RF to both in the deuterated Ultrapure and Control systems.

We are aware of that it is possible to obtain $V_{RDeE}$ scale from other reference electrode potentials, such as V vs Ag/AgCl or vs Hg/HgSO$_4$.[55] However, a RDeE is recommended to use as a reference electrode instead of these typical rones. This is because the method to use a RHE for heavy water systems can be applied in a relatively limited acidic concentration range, and also the use of a RDeE is free from suffering from unknown effects, such as liquid junction potential, therefore it is more accurate approach than a use of other reference electrodes.

**2.2. Analytical Methods for Kinetics of Electrode Processes**

The electron transfer number ($n$) and D$_2$O$_2$ yield were determined by the following equations:

$n = 4*I_d/(I_d+I_r/N)\}$,

D$_2$O$_2$ yield (%) = $200*I_r/N/(I_d + I_r/N)$,

D$_2$O yield (%) = 100 (%) − D$_2$O$_2$ yield (%),

where $I_d$ is disc current, $I_r$ is ring current and $N$ is a collection efficiency which was determined to be 0.39 in the systems.

Transfer coefficient $\alpha$ can be obtained from Tafel slope $b$ by using Eq. 1, where $R$ and $T$ are the gas constant and the temperature (298 K in this study), respectively.

$$b = \frac{2.303RT}{\alpha F} \qquad \text{Eq. 1}$$

In order to observe kinetic isotope effect (KIE), a KIE rate constant ratio ($k^H/k^D \equiv K^{H/D}$) was obtained in each system. The superscripts H and D shows a rate constant in the ordinary and heavy water systems, respectively. Based on the previous reports, a $K^{H/D}$ of ORR can be obtained by the following equation:

$$K^{H/D} = \frac{k_0^H}{k_0^D} = \frac{j_0^H}{j_0^D} \times \frac{C_0^D}{C_0^H} \times \exp\left\{\frac{(\alpha^D - \alpha^H)F\eta}{RT}\right\}, \qquad \text{Eq. 3}$$

where $j_0$, $k_0$, $C_0$, $\eta$, and $F$ are, an exchange current density, a rate constant for $j_0$, oxygen concentration, overpotential, and Faraday constant, respectively. The $C_0^D/C_0^H$ is known to be 1.101.[35]

Due to the following equation, we can obtain $j_0$ from a corresponding log $j_k$-$\eta$ diagram:

$$j_k = j_0 \times exp\left\{-\frac{\alpha F}{RT}\eta\right\}$$

$$\Leftrightarrow \log j_k = \log j_0 - \frac{\alpha F}{2.303RT}\eta$$

$$\Leftrightarrow \eta = -b\,(\log j_k + \log j_0) \qquad \text{Eq. 4}$$

The ORR kinetic currents on Pt electrode can be separated from diffusion limiting current by using a following equation since we can observe clear diffusion-limited currents:

$$\frac{1}{j} = \frac{1}{j_k} + \frac{1}{j_{lim}} \Leftrightarrow j_k = \frac{(j_{lim} \cdot j)}{(j_{lim} - j)},$$

where $j$ and $j_{lim}$ are experimentally obtained current with RRDE technique and diffusion limiting current, respectively.

The equilibrium potential for $D_2O$ formation ($E^0_{D2O}$) is 1.262 $V_{RDeE}$,[32] therefore the overpotential for ORR in heavy water systems is

$\eta$ (V) = Experimentally obtained potential ($V_{RDeE}$) − 1.262 ($V_{RDeE}$).

## 3. RESULTS AND DISCUSSION

### 3.1. Electrochemistry in Fully Deuterated Acidic Electrolyte

### 3.1.1. Ion Adsorption/Desorption onto Pt Surface in Deuterated Acidic Solution

The cyclic voltammograms (CV) of polyPt were obtained in 0.05M $D_2SO_4$ in Ultrapure and Control $D_2O$, respectively (Figure 1a). As it is obvious from this figure, the Control system does not show appropriate electric double layer (EDL) region from 0.3 to 0.7 $V_{RDeE}$ in case of a positive scan direction, however, show a shallow increase of anodic current starting from around 0.4 $V_{RDeE}$. Furthermore, the anodic peak in the Ultrapure starting from 0.8 $V_{RDeE}$, suggesting $OD^-$ adsorption and following surface oxide formation, shifts in case of the Control. Indeed, the shapes due to the deuteron adsorption/desorption (D-UPD) currents are different between the two systems: the D-UPD peaks in the Ultrapure shows much pronounced peaks. Furthermore, if we measure the $D^+$ desorption isotherm for the Ultrapure, it is identical to $H^+$ one (Figure S1). These results indicate that, although we used a highly pure distilled $D_2O$, the Control contains impurities. However, beside the different $D^+/H^+$ adsorption/desorption behavior, the cathodic peak position in Ultrapure corresponding to surface oxide reduction (0.78 $V_{RDeE}$) is identical to the peak position in Control. In order to check the surface condition of the Pt electrode, we measured the isotherm, which shows a charge corresponding to the deuteron under potential deposition (D-UPD), denoted as $Q_{D-UPD}$ (Figure 1b). The $Q_{D-UPD}$ differs in the two systems. In Ultrapure system, $Q_{D-UPD}$ is 220 µC cm$^{-2}_{ECSA}$ but 191 µC cm$^{-2}_{ECSA}$ in Control system, even with the same polyPt electrode. We note that the ECSA of the electrode was obtained by using the ultrapure ordinary electrolyte, and used the same electrode in the deuterated systems (See the experimental section). Based on a huge amount of discussions on the electrolyte contamination, the impurity can be organic compounds, and this can be a reason for a larger anodic current of Control from 1.0 to 1.5 V than that of Ultrapure, and also the decrease of $Q_{D-UPD}$ in Control system. In addition to this, the observation of current at EDL region in Control is also a strong evidence of impurity (Figure 1a).

**3.1.2. ORR Electrode Processes of Pt Surface in Fully Deuterated Acidic Solution**
The ORR kinetics in the acidic condition were investigated by using the polyPt electrode to check the effect of the different electrolytes toward the ORR activities. All electrochemical properties in the acidic condition are summarized in the Table 1. First of all, linear sweep voltammetry (LSV) with RRDE technique was applied to check the ORR kinetics, and surprisingly, we found that there is only a small effect of impurities (Figure 2a). In addition to this result, the two systems show $n = 3.98$ for a potential range from 0.2 to 0.9 $V_{RDeE}$ (Figure 2b), suggesting the main product is heavy water. The four-electron ORR pathway of polyPt in acidic condition is well agreed with previous reports using ordinary water, therefore, we conclude that the ORR electrode process undergoes property in the two deuterated electrolytes.[10]

In order to acquire further insights of the electrode process, the log $j_k$-$\eta$ diagrams were analyzed (Figure 3a). The Tafel relation is applied to analyze the diagram. This means that we hypothesized the existence of at least one linear relation between log $j_k$ and $\eta$ in a certain range of potential. In this study, two Tafel lines are taken to analyze the kinetics of ORR on polyPt in acidic conditions, therefore we fitted the diagram with lines in two regions, i.e. $\eta > 0.35$ V (the corresponding Tafel slope (TS) is denoted as $TS_1$) and < 0.40 V (denoted as $TS_2$). This is because polyPt in acidic condition usually does not show a clear linear region in the log $j_k$-$\eta$ diagram of ORR, and it was often reported that there are the two lines for this condition.[5, 56]

The $TS^D_1$ in the Ultrapure was 68.6 mV/dec, which is almost identical to the $TS^H_1$ in the ultrapure ordinary electrolyte obtained as 69.1 mV/dec (Table 1). From these values, we obtained log $j_0$ in the ultrapure deuterated and ordinary systems as −8.53 and −8.37, respectively. Furthermore, we set $\alpha^H = \alpha^D$ for Eq. 3 owing to the same values of $TS_1$ in these conditions. This leads to $K^{H/D}_1$ in Ultrapure as 1.59. If we assume that $\alpha^H \neq \alpha^D$ in case of $TS^H_1 \neq TS^D_1$, Eq. 3. becomes a potential-dependent equation, and we obtain $K^{H/D}_1$ = 1.47 at $\eta = -0.325$. Therefore, there is only a tiny difference between the case for $TS^H_1 = TS^D_1$ and $TS^H_1 \neq TS^D_1$. With the same approach, we obtained $TS^D_2$ in Ultrapure and $TS^H{}_2$ as 100 and 99.3 mV/dec, respectively. Based on this result, we can assume $TS^H{}_2 = TS^D{}_2$, and this leads to $K^{H/D}{}_2 = 1.26$. Both $K^{H/D}$ values are close to the one obtained by Yeager and his co-workers ($K^{H/D} = 1.4$).[32]

Moving to the Control, the $TS^D{}_1$ in this condition was 71.6 mV/dec (Table 1). From this value, we obtained log $j_0$ in the Control as −8.34. Due to this difference in TS, we set $\alpha^H \neq \alpha^D$, therefore we obtained $\alpha^H{}_1 = 0.86$ and $\alpha^D{}_1 = 0.84$ from Eq. 1, and applied these values to Eq. 3. This leads to $K^{H/D}{}_1$ in the Control as 1.43 at $\eta = -0.325$. With the same approach, we obtained $\alpha^H{}_2 = 0.60$ and $\alpha^D{}_2 = 0.57$ and therefore $K^{H/D}{}_2$ in Control = 1.75 at $\eta = -0.425$. As it is indicated by the above data, the KIE on ORR in the acidic condition is less influenced by impurities even though the clear differences were observed in CV and LSV (Figure 1 and Figure 2a).

### 3.2. Electrochemistry in Fully Deuterated Alkaline Electrolyte

### 3.2.1. Ion Adsorption/Desorption onto Pt Surface in Deuterated Alkaline Solution

In Figure 4a, it was shown the CV of polyPt in 0.1M KOD in Ultrapure and Control $D_2O$. These results show huge differences of the electrode processes in the Ultrapure and Control: (1) The peaks corresponding to D-UPD in Control are shifted and weakened compared to the Ultrapure; and (2) the anodic peaks usually corresponding to $OD^-$ adsorption and oxide formation in the Control are much more pronounced and negatively shifted compared to the Ultrapure; (3) The $D^+$ desorption isotherms in the deuterated and protonated Ultrapure suggest that the deuterated Ultrapure is as clean as the protonated Ultrapure, which is well-established to prepare a highly clean electrochemical system (Figure S2).

Furthermore, we compared the isotherm corresponding to $Q_{\text{D-UPD}}$ for the Ultrapure with the Control in order to check the surface conditions (Figure 4b). The isotherms differs in the two systems. In the Ultrapure system, $Q_{\text{D-UPD}}$ is 220 µC cm$^{-2}_{\text{ECSA}}$ but 157 µC cm$^{-2}_{\text{ECSA}}$ in the Control system, even with the same polyPt electrode. This huge difference of $Q_{\text{D-UPD}}$ between the Ultrapure and Control suggest that impurity gives a stronger effect to alkaline conditions that to acidic conditions, and this can be a reason for a larger anodic current of Control from 0.6 to 1.5 V than that of Ultrapure, and also the decrease of $Q_{\text{D-UPD}}$ in Control system, which observations are same for the Control in the acidic condition.

### 3.2.2. ORR Electrode Processes on Pt Surface in Fully Deuterated Alkaline Solution

The ORR electrode processes in the alkaline condition were investigated to check the effect of the different electrolytes toward the ORR activities. All electrochemical properties in the alkaline conditions are summarized in the Table 2. First of all, LSV curve was shown to check the ORR kinetics. As the results, on the contrary to the acidic condition, we found that there are significant effects owing to different electrolytes. The LSV curve of the Control much negatively shifted compared to the Ultrapure (Figure 5a). This shows that the catalytic activity of Pt was obviously altered because of the different electrolytes. In addition to these results, beside it was constantly shows $n = 3.98$ in the Ultrapure in a potential range from 0.2 to 0.9 V$_{\text{RDeE}}$, in the Control, $n$ shows three plateaus, *i.e.* $n =$ below 3.95, 3.96, and 3.98 in the potential ranges from 0.7 to 0.9 V$_{\text{RDeE}}$, 0.4 to 0.6 V$_{\text{RDeE}}$, and < 0.4 V$_{\text{RDeE}}$ respectively (Figure 5b). The $n$ value in the potential ranges from 0.4 to 0.9 V$_{\text{RDeE}}$ is lower than typical values reported in previous studies.[57] Furthermore, it is unusual to observe several plateaus in $n$ value. These results indicate that although the main product is heavy water, however a higher amount of deuteron peroxide was produced in case of the Control compared to the Ultrapure case. Therefore, the two-electron pathway as the side reaction was enhanced in the Control. In addition to this, the reaction path could be altered on changing electrode potential as evident by a change of $n$ value on applied overpotential (Figure 5b).

The log $j_k$-$\eta$ diagrams were studied in order to acquire further insights of the electrode process (Figure 6). In the alkaline conditions, only one Tafel line was observed in each condition. The TS$^D$ in the Ultrapure was 47.6 mV/dec, which is almost identical to the TS$^H$ in the ultrapure ordinary electrolyte obtained as 47.7 mV/dec (Table 2). From these values, we obtained log $j_0$ in the ultrapure deuterated and ordinary systems as −9.92 and −9.17, respectively. This leads to $K^{H/D}$ in the Ultrapure as 6.19 since $\alpha^H = \alpha^D$ for Eq. 3. A higher $K^{H/D}$ value compared to the one in the acidic condition could be related to the different microscopic proton transfer mechanism. Nevertheless, this $K^{H/D}$ value is same for the result obtained in the previous report,[37] and this can be explained by the semiclassical transition state theory.[58] From this point of view, $K^{H/D}$ = 6.19 indicates that a trivial proton transfer process, might be combined with a tunneling correction,[1, 59] is related to the rate-determining step of the corresponding reaction.

However, the Control system showed the quite different feature. The TS$^D$ in the Control was 88.8 mV/dec (Table 2). From this value, we obtained log $j_0$ in the Control as −7.77. Due to a large difference of TS in the Ultrapure and the Control, the Eq. 3 in this condition becomes a potential-dependent equation, and a large difference of TS leads the $K^{H/D}$ value 37 at $\eta$ = −0.3 and more than 1×10$^3$ at $\eta$ > −0.45. Of course it is possible to conclude that a high $K^{H/D}$ value is a result of quantum tunneling dominating the proton transfer in RDS if the corresponding system is enough clean and therefore experiment is reliable.[58] However, this high $K^{H/D}$ value in the Control is not related to these non-trivial effects but because of impurity as indicated by the CV and isotherms (Figure 5). As such, these results show that an ultrapure condition is indispensable for reliable measurements of KIE, especially in alkaline conditions.

## 4. CONCLUSION AND OUTLOOK

Aiming to establish a reliable approach to measure KIE in electrocatalysis, we have studied the ORR kinetics and the corresponding KIE by using different purity of fully deuterated acidic and alkaline electrolytes. The main conclusions are following:

(1) Although the obvious difference in of CV and LSV, there is a less impact to use ultrapure deuterated electrolyte to measure KIE in acidic conditions. The $K^{H/D}$ values are similar for the Control and the Ultrapure. On the other hand, in case of alkaline conditions, a use of ultrapure deuterated electrolytes is indispensable to measure reliable KIE and make reasonable discussions. In case of the Control in alkaline conditions, $TS^D$ was quite different compared to $TS^H$ because of impurity in the electrolyte, and this led to a high $K^{H/D}$ value of $> 1\times10^3$ at $\eta > -0.45$, however $K^{H/D} = 6.19$ in case of the Ultrapure, which value is as same as the previous report.

(2) The impurity mainly comes from the as-received $D_2O$. We can solve this problem with a commercially available equipment to obtain ultrapure quality $D_2O$. This is exactly same previous efforts to obtain high quality ordinary water for an ultraclean therefore a reliable electrochemical system.

(3) In order to prepare an equivalent level of purity in deuterated systems as clean as protonated systems with Milli-Q water, ultrapure deuterated electrolytes should be used. With this ultrapure electrolyte, we can obtain reliable KIE in electrocatalysis combined with appropriate analytical equations and well-established approaches to prepare a highly clean electrochemical system.

All-embracing, this study shows that ultrapure deuterated electrolyte is a key to measure reliable KIE in multi-electron/-proton transfer electrode processes. The approach shown here can be applied to a wide spectrum of research on electrode/electrolyte interfaces combining with surface spectrometry or first-principle calculations in order to unveil complicated reactions at electrode/electrolyte interfaces.

**Acknowledgement**

This study was partially supported by JSPS KAKENHI 17K14546 and 19K15527. KS is indebted to NIMS and Japan Prize Foundation.

18

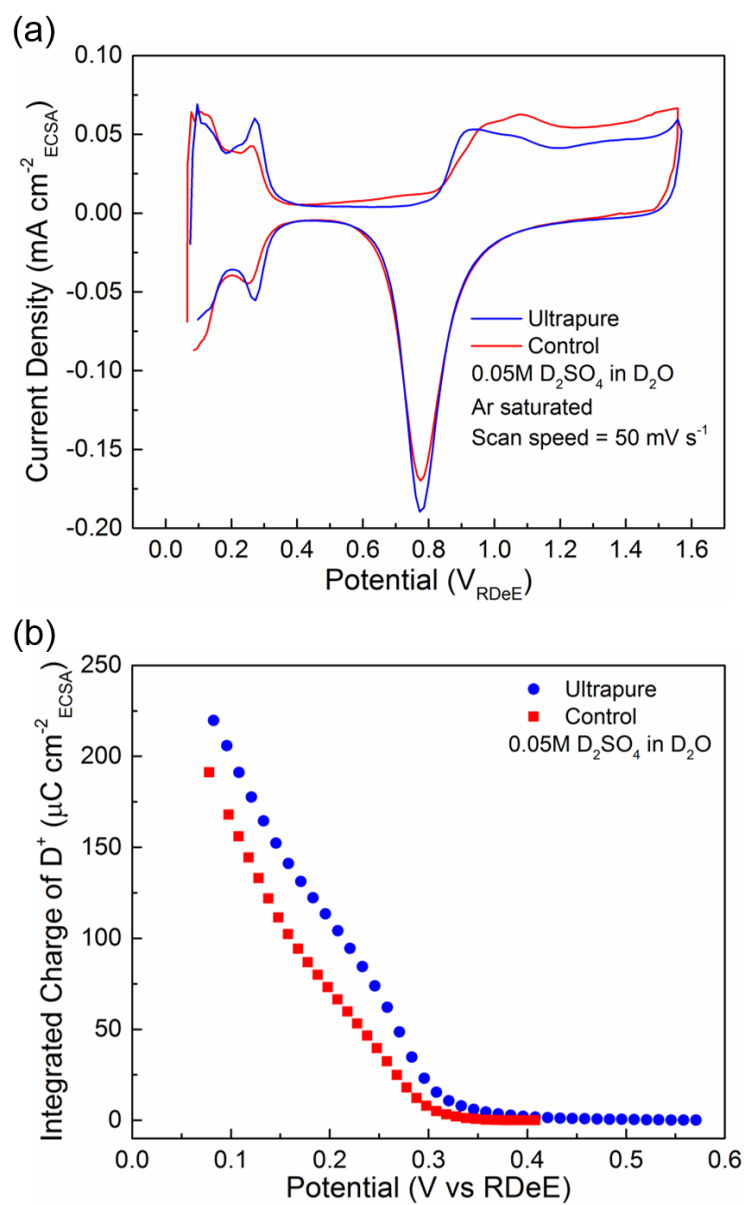

**Figure 1. Fundamental electrochemical analysis on pcPt electrode in fully deuterated acidic electrolytes (0.05M $D_2SO_4$ in $D_2O$)**. (a) Cyclic voltammograms in the Ultrapure (blue solid line) and Control (red solid line) electrolytes, and (b) $D^+$ desorption isotherm. Plots corresponding to D-UPD in the Ultrapure the Control electrolytes are shown by blue filled circles and red filled squares, respectively.

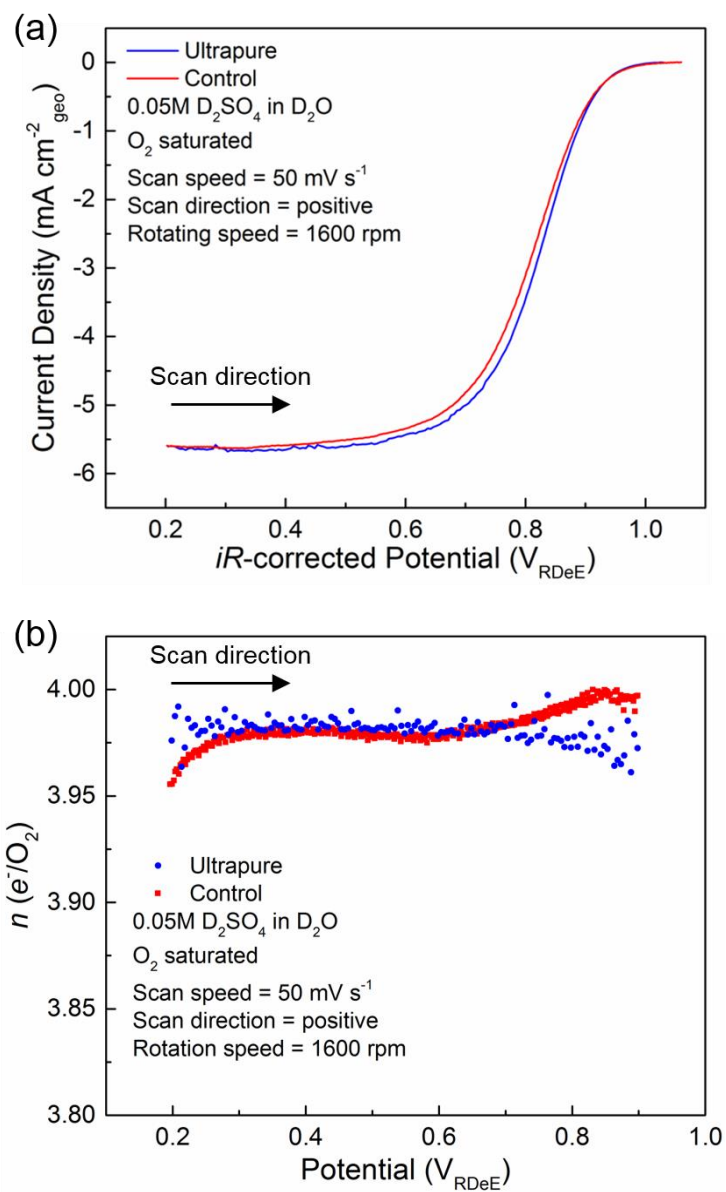

**Figure 2. ORR on pcPt electrode in fully deuterated acidic electrolytes (0.05M $D_2SO_4$ in $D_2O$).** (a) LSV in the Ultrapure (blue solid line) and Control (red solid line) electrolytes, and (b) Electron transfer number to dioxygen molecule ($n = e^-/O_2$) on changing potential in the Ultrapure (blue filled circles) and Control (red filled squares) electrolytes.

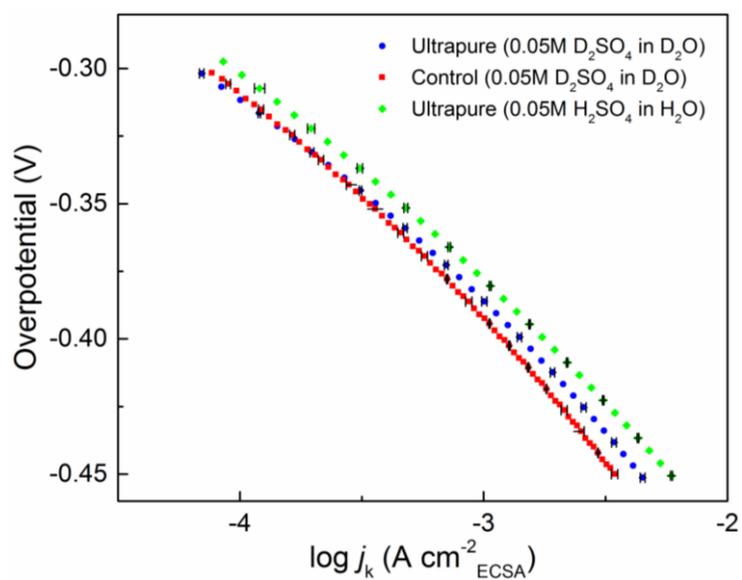

**Figure 3. Log $j_k$ vs Overpotential diagram in acidic electrolytes.** The measurements in the Ultrapure and Control fully deuterated acidic electrolytes (0.05M $D_2SO_4$ in $D_2O$) are indicated as blue filled circles and, respectively. The measurement in the Ultrapure ordinary acidic electrolyte (0.05M $H_2SO_4$ in $H_2O$) is indicated as green filled diamonds. Error bars are shown in each three points.

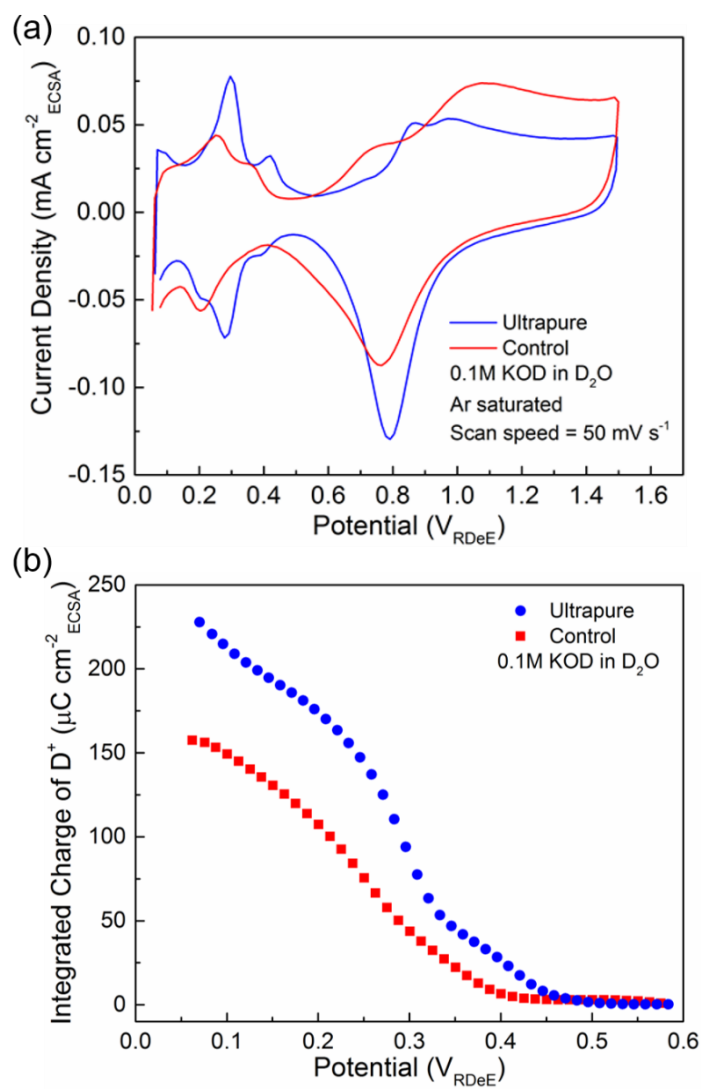

**Figure 4. Fundamental electrochemical analysis on pcPt electrode in fully deuterated alkaline electrolytes (0.1M KOD in D$_2$O)**. (a) Cyclic voltammograms in the Ultrapure (blue solid line) and Control (red solid line) electrolytes, and (b) D$^+$ desorption isotherm. Plots corresponding to D-UPD in the Ultrapure the Control electrolytes are shown by blue filled circles and red filled squares, respectively.

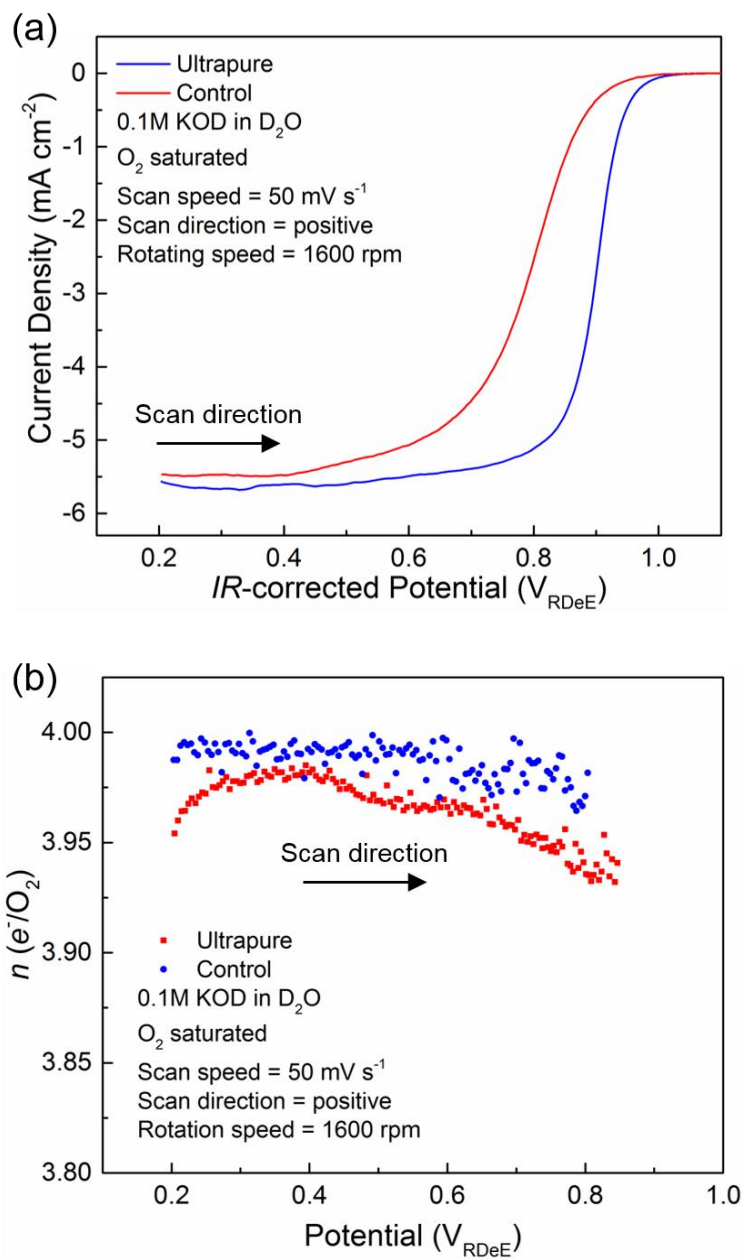

**Figure 5. ORR on pcPt electrode in fully deuterated alkaline electrolytes (0.1M KOD in D$_2$O).** (a) LSV in the Ultrapure (blue solid line) and Control (red solid line) electrolytes, and (b) Electron transfer number to dioxygen molecule ($n = e^-/O_2$) on changing potential in the Ultrapure (blue filled circles) and Control (red filled squares) electrolytes.

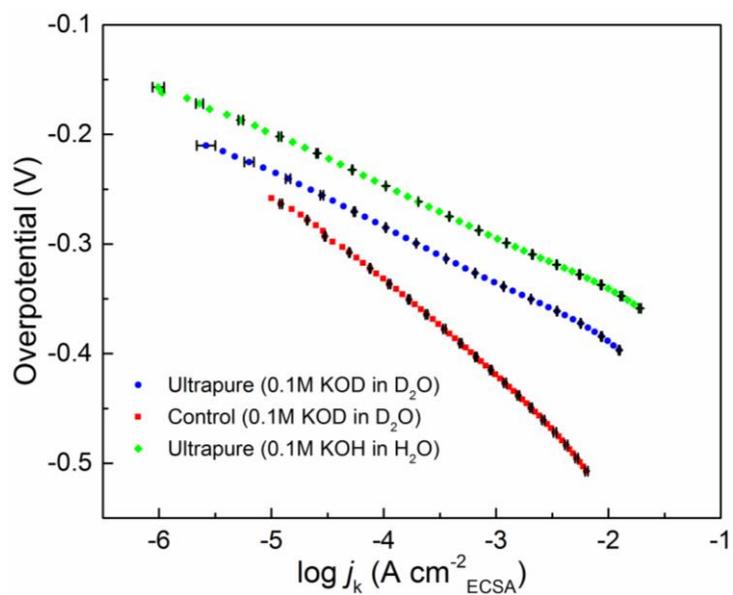

**Figure 6. Log $j_k$ vs Overpotential diagram in alkaline electrolytes.** The measurements in the Ultrapure and Control fully deuterated alkaline electrolytes (0.1M KOD in $D_2O$) are indicated as blue filled circles and, respectively. The measurement in the Ultrapure ordinary alkaline electrolyte (0.1M KOH in $H_2O$) is indicated as green filled diamonds. Error bars are shown in each three points.

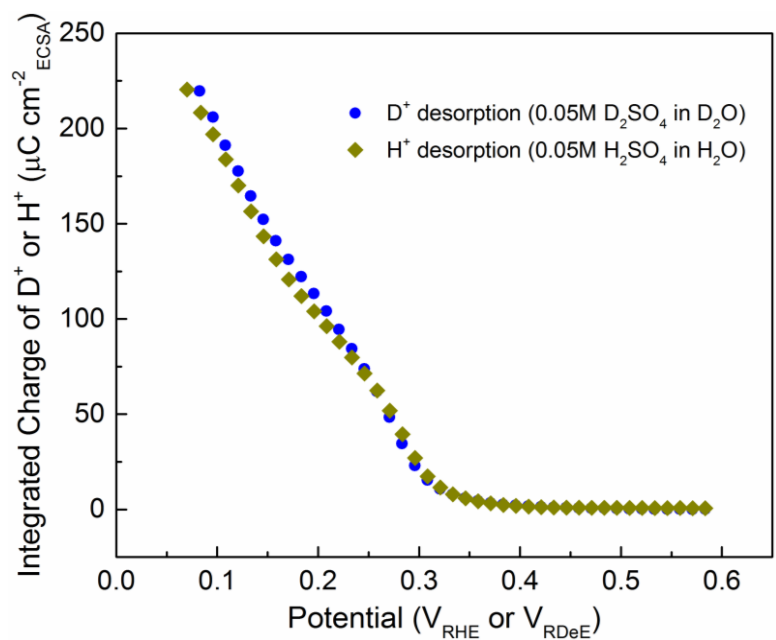

**Figure S1.** $D^+/H^+$ desorption isotherms in the deuterated and protonated ultrapure acidic conditions.

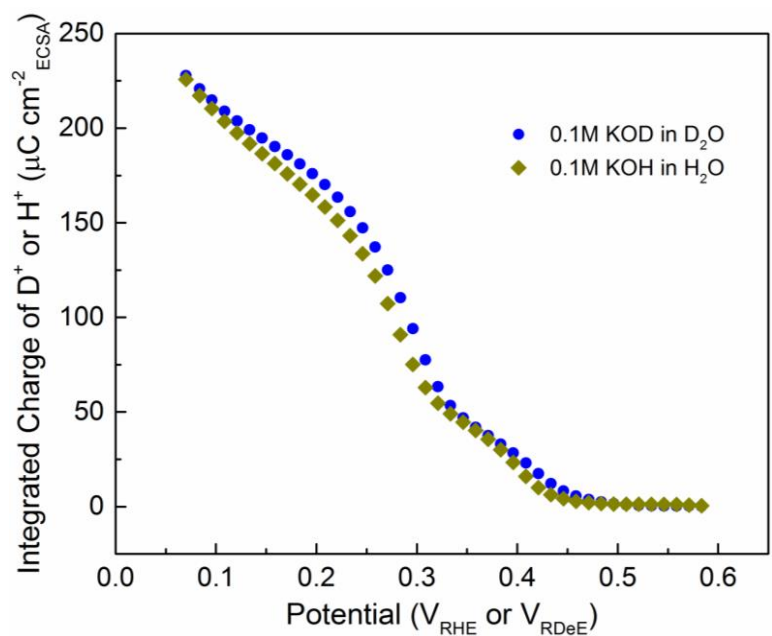

**Figure S2.** $D^+/H^+$ desorption isotherms in the deuterated and protonated ultrapure alkaline conditions.

**Table 1.** ORR kinetic values in the deuterated and protonated acidic electrolytes.

| Electrolyte | $TS_1$ (mV/dec) | $TS_2$ (mV/dec) | $-\log j_{0\_1}$ (A/cm$^2_{ECSA}$) | $-\log j_{0\_2}$ (A/cm$^2_{ECSA}$) | $K^{H/D}_1$ | $K^{H/D}_2$ |
|---|---|---|---|---|---|---|
| Ultrapure | 68.6 ± 0.7 | 100 ± 1 | 8.53 | 6.82 | 1.59 | 1.26 |
| Control | 71.2 ± 0.6 | 104 ± 1 | 8.34 | 6.75 | 1.43[a] | 1.75[b] |
| Ultrapure (Ordinary) | 69.1 ± 0.5 | 99.3 ± 0.6 | 8.37 | 6.76 | - | - |

[a]This value was obtained at $\eta = -0.3$. [b]This value was obtained at $\eta = -0.45$.

**Table 2.** ORR kinetic values in the deuterated and protonated alkaline electrolytes.

| Electrolyte | TS (mV/dec) | $-\log j_0$ (A/cm$^2_{ECSA}$) | $K^{H/D}$ |
|---|---|---|---|
| Ultrapure | 47.6 ± 0.8 | 9.92 | 6.19 |
| Control | 88.8 ± 0.8 | 7.77 | 37 [a], >1×10$^3$ [b] |
| Ultrapure (Ordinary) | 47.7 ± 0.2 | 9.17 | - |

[a]This value was obtained at $\eta = -0.325$. [b]This value was obtained at $\eta = -0.425$.